\documentclass[11pt]{article}

\usepackage{amsmath,amsthm,amsfonts}
\usepackage[round]{natbib}
\RequirePackage[dvips]{graphicx}
\usepackage{lscape,graphicx,rotating,rotfloat,multirow,booktabs,ctable,threeparttable}

\theoremstyle{definition}

\newcommand{\btheta}{\mbox{\boldmath $\theta$}}

\newcommand{\bxi}{\mbox{\boldmath $\xi$}}

\newcommand{\bmu}{\mbox{\boldmath $\mu$}}

\newcommand{\bG}{\mbox{\boldmath $G$}}

\newcommand{\br}{\mbox{\boldmath $r$}}
\newcommand{\bs}{\mbox{\boldmath $s$}}
\newcommand{\bz}{\mbox{\boldmath $z$}}
\newcommand{\by}{\mbox{\boldmath $y$}}
\newcommand{\bh}{\mbox{\boldmath $h$}}
\newcommand{\bInf}{\mbox{\boldmath $I$}}

\newenvironment{double}{\Large
\normalsize}{\Large\normalsize}

\newcommand{\blankfoot}[1]%
{{%
\footnotetext{#1}} }

\begin{document}

\pagestyle{plain} \vspace{1in}

\title{\LARGE \bf Riemann Manifold Langevin Methods on Stochastic Volatility Estimation}

\author{
  Mauricio Zevallos\footnote{Department of
  Statistics, University of Campinas (UNICAMP), Brazil. E-mail:
  {\texttt amadeus@ime.unicamp.br}} 
  \and Loretta Gasco\footnote{Pontificia Universidad Cat\'olica del
  Per\'u. E-mail: lgasco@pucp.edu.pe}
  \and Ricardo Ehlers\footnote{Corresponding author. University of S\~ao Paulo,
  Brazil. E-mail: ehlers@icmc.usp.br}
}

\date{\today}

\maketitle

\begin{double}

\begin{abstract}
In this paper we perform Bayesian estimation of stochastic volatility
models with heavy tail distributions using Metropolis adjusted
Langevin (MALA) and Riemman manifold Langevin (MMALA) methods. We
provide analytical expressions for the application of these methods,
assess 
the performance of these methodologies in simulated data and illustrate
their use on two financial time series data sets.\\ 

\noindent {\bf Keywords:} Bayesian, Markov chain Monte Carlo, Metropolis-Hastings, Value at Risk.
\end{abstract}

\section{Introduction}

Stochastic volatility (SV) models were proposed by \cite{taylor86}. This
model and its generalizations has been applied successfully to model
the time-varying volatility present in financial time series. To
estimate these models several estimation methods 
have been proposed in the literature, quasi-maximum likelihood methods
\citep{har94}, generalized method of moments \citep{and-sor96},
Markov Chain Monte Carlo Methods (MCMC) \citep[pioneered by][]{jpr94}
and Integrated Nested Laplace Approximations \citep{martino-etal10},
to name a few. For an account of recent
developments in the estimation of SV models see 
\cite{broto-ruiz04} and \cite{shep-and09}
and the references therein.
In particular, MCMC methods are considered one of the most efficient estimation
method. Proposals include for example \cite{jpr94} and \cite{kim98}.

Recently, \cite{giro11} proposed a methodology based on Metropolis
adjusted Langevin and Hamiltonian Monte Carlo sampling methods. 
These methods take advantage of the relationship between Riemann
geometry and statistics to overcome some of the shortcomings of
existing Monte Carlo algorithms. They provide 
evidence that some sort of local calibration in the MCMC scheme may
lead to strong improvements in large dimensional problems.

In particular, one of the examples discussed by these
authors is the estimation of SV models with normal
perturbations. Since these models often give rise to posterior
distributions with high correlations the methods proposed can be
particularly useful for estimation. More recently,
\cite{nugroho-morimoto} presented an algorithm based on Hamiltonian Monte Carlo
methods for the estimation of realized stochastic volatility models.

In this paper we discuss the use Langevin and Modified Langevin methods to the
estimation of SV models with $t$-Student and GED perturbations for the
observations. We give the expressions, assess the performance and
illustrate with two real data sets. 

Because the computational time is
critical for stochastic volatility models we implemented a hybrid
method in which a Riemann manifold MALA (MMALA) scheme is applied for
the parameters and a MALA scheme is applied for the volatilities.
In particular, all the computations in this paper were implemented
using the open-source statistical software language and environment {\tt R}
(\cite{R06}).

The remainder of this paper is organized as follows. The models are
presented in Section \ref{models} and the methodology for estimation
is discussed in Section \ref{estimation}. To assess the estimation
methodology some Monte Carlo experiments are presented in Section
\ref{simulations}.  Section \ref{illustrations} illustrates with
empirical data, and 
some final remarks are given in Section \ref{conclusions}.

\section{Models}  \label{models}

We consider the following Stochastic Volatility (SV) model, 
\begin{eqnarray}
y_t & = & \beta\exp(h_t/2) \varepsilon_t, \label{sv-eqn1}\\
h_t & = & \phi h_{t-1} + \eta_t, \label{sv-eqn2}
\end{eqnarray}
where $\{\varepsilon_t\}$ is a sequence of independent identically
distributed (IID) random variables with zero mean and unit variance,
$\{\eta_t\}$ is an IID sequence of random variables such that 
$\eta_t \sim N(0,\sigma^2)$,  $\eta_t$ and $\varepsilon_t$ are
independent for all $t$. In addition, we assume that $\beta > 0$ and
$|\phi| < 1$.

In the SV model, conditional to the information set
$\mathcal{F}_t =\{y_t, y_{t-1},\ldots\}$, the standard deviation of
$y_t$ is given by,
\begin{eqnarray*}
\sigma_t = \beta \exp(h_t/2).
\end{eqnarray*}
In Finance, if $y_t$ represents the $t$-th return then $\sigma_t$  is
the {\it volatility} at time $t$.  

The original formulation of the SV model by \cite{taylor86} considers
$\varepsilon_t$ following a standard normal  
distribution.  However, many empirical studies indicate that this
model does not account for the kurtosis 
observed in most financial time series returns. Consequently, several other
error distributions have been considered. 
For example, we consider  $\varepsilon_t$ following an Exponential
Power distribution (or generalized error distribution, GED)
with zero mean, unit variance 
(see \cite{boxtiao73} and \cite{nelson91}) with density function,
\begin{eqnarray} \label{ged}
f(\varepsilon_t)  =  \frac{\nu}{\lambda 2^{1+1/\nu}\Gamma(1/\nu)} \exp
\left \{-\frac{1}{2}\left| \frac{\varepsilon_t}{\lambda} \right|^\nu
\right\} 
\end{eqnarray}
where $\lambda^2= 2^{-2/\nu}\Gamma(1/\nu)/\Gamma(3/\nu)$ and the shape
parameter $\nu >0$.  
Important special cases are, the Laplace (or double exponential)
distribution for $\nu=1$ and the standard normal distribution when $\nu=2$.
The kurtosis is given by 
$\Gamma(1/\nu) \Gamma(5/\nu)/\Gamma(3/\nu)^2 -3$ 
so that when $\nu < 2$ this distribution reproduces heavy-tails. In
addition, we consider $\varepsilon_t$ following a $t$-Student
distribution with $\nu$ degrees of freedom and density function, 
\begin{eqnarray} \label{Student}
f(\varepsilon_t)  =  \frac{1}{\sqrt{\pi (\nu-2)}}
\frac{\Gamma(\frac{\nu+1}{2})}{\Gamma(\frac{\nu}{2})}  
\left \{1 + \frac{\varepsilon_t^2}{\nu-2} \right\}^{-(\nu+1)/2}.
\end{eqnarray}
When $\nu \to \infty$ this distribution approaches the standard
normal distribution. 

\section{Estimation} \label{estimation}

Let $y_1,\ldots,y_n$ be the observed time series. In order to estimate
this model we use the Metropolis adjusted Langevin (MALA) and the
Riemannian Manifold Metropolis adjusted Langevin (MMALA) Monte Carlo methods
proposed by \cite{giro11}. The estimation
procedure is performed in a two-step blocking approach. In the first
step, the latent variables $\{h_t\}$ (the log-squared volatilities)
are sampled and then, conditional on these sampled values, we sample
the parameters $\btheta = (\beta,\sigma,\phi,\nu)$. At each step, a
Metropolis-Hastings sampling scheme is applied using the methods described
below. 

\subsection{Metropolis adjusted Langevin algorithm (MALA)}

Let $\bxi\in\mathbb{R}^D$ be the random vector of interest with
density $f(\bxi)$. Then the Metropolis adjusted Langevin algorithm
MALA is based on a Langevin diffusion process whose stationary
distribution is $f(\bxi)$ and its stochastic differential equation is
discretized to give the following proposal mechanism,
\begin{eqnarray}\label{eq-mala-1}
\bxi = 
\bxi^{[n]} + \frac{\epsilon^2}{2}\nabla_{\xi} \ln f(\bxi^{[n]}) + \epsilon\bz 
\end{eqnarray}
where $\bz \sim N(0,\bInf)$ with $\bInf$ the identity matrix of
order $D$ and $\epsilon$ is
the integration step size. A Metropolis acceptance probability is then
employed to ensure convergence to the invariant distribution as
follows. A new 
value $\bxi$ is sampled from a multivariate normal distribution with
mean $\mu(\bxi^{[n]},\epsilon) = \bxi^{[n]} + \frac{\epsilon^2}{2}
\nabla_{\xi} \ln f(\bxi^{[n]})$ and variance-covariance matrix
$\epsilon^2\bInf$. This value is accepted with probability given by
$\min \{1,
f(\bxi) q(\bxi^{[n]} | \bxi)/ f(\bxi^{[n]}) q(\bxi | \bxi^{[n]} )
\}$ where the proposal density is 
$q(\bxi | \bxi^{[n]} ) = N(\mu(\bxi^{[n]},\epsilon),\epsilon^2 \bInf)$.

This algorithm is then employed to estimate the SV model following the two
steps below.

\begin{enumerate}

\item {\it Sample the latent variables} $\bh$. Assuming the parameters
  as constants, apply (\ref{eq-mala-1}) with $f = f(\by,\bh)$ and
  gradient $\nabla$ calculated with respect to $\bh$. 

\item {\it Sample parameters} $\btheta$. Given $(\by,\bh)$, apply
  (\ref{eq-mala-1}) with $f=f(\by,\bh |\btheta)f(\btheta)$ and
  gradient $\nabla$ calculated with respect to $\btheta$.

\end{enumerate}

\subsection{Riemann Manifold MALA (MMALA)}

\cite{giro11} developed a modification in the Metropolis proposal
mechanism in which the moves in $\mathbb{R}^D$ are according to a
Riemann metric instead of the standard Euclidian distance.
This procedure is refered to as Riemann manifold MALA or
MMALA. The proposal mechanism is now given by,
\begin{eqnarray}
\xi_i & = & \mu(\bxi^{[n]},\epsilon)_{i} + \left\{\epsilon
\sqrt{\bG^{-1}}(\bxi^{[n]})\bz \right\}_{i}, \label{eq-mmala-1}\\ 
\mu(\bxi^{[n]},\epsilon)_{i} & = & \xi^{[n]}_{i} +
\frac{\epsilon^2}{2} \left\{ \bG^{-1}(\bxi^{[n]})\nabla_{\xi} \ln
f(\bxi^{[n]}) \right\}_{i} \nonumber \\
&-& \epsilon^2 \sum_{j=1}^D \left\{ \bG^{-1}(\bxi^{[n]})   \frac{d
  \bG(\bxi^{[n]})}{d\xi_j} \bG^{-1}(\bxi^{[n]})
\right\}_{ij}\nonumber \\ 
& + & \frac{\epsilon^2}{2} \sum_{j=1}^D \left\{ \bG^{-1}(\bxi^{[n]})
\right\}_{ij} tr\left\{\bG^{-1}(\bxi^{[n]})\frac{d
  \bG(\bxi^{[n]})}{d\xi_j}\right\}\label{eq-mmala-2}  
\end{eqnarray}
where $\bz\sim N(0,\bInf)$ and,
\begin{eqnarray*} \label{eq-mmala-3}
\bG(\bxi) = -E\left(\frac{d^2 \ln f(\bxi)}{d\bxi^\top\bxi}  \right).
\end{eqnarray*}

\noindent Then, employing a Metropolis mechanism with
proposal density given by $q(\bxi | \bxi^{[n]} ) =
N(\mu(\bxi^{[n]},\epsilon),\epsilon^2  \bG^{-1}(\bxi^{[n]}) )$ and the usual
acceptance probability given by the quantity
$\min \{1,f(\bxi) q(\bxi^{[n]} |\bxi)/f(\bxi^{[n]}) q(\bxi|\bxi^{[n]} )\}$ 
ensures convergence to the invariant distribution. We note that in
this case both the mean vector and covariance matrix in the proposal
distribution depend on the current state of the Markov chain.

A simplified proposal mechanism is obtained when a constant curvature is
assumed. In this case, the last two terms in (\ref{eq-mmala-2}) vanish
and the proposal mean becomes,
\begin{eqnarray*}
\bmu(\bxi^{[n]},\epsilon) = \bxi^{[n]} + \frac{\epsilon^2}{2}
\bG^{-1}(\bxi^{[n]})\nabla_{\xi} \ln f(\bxi^{[n]}).
\end{eqnarray*}
In this simplified version of MMALA, the state-dependent covariance
matrix in the proposal mechanism 
still allows adaptation to the local curvature of the target $f(\bxi)$ which
has been shown to increase algorithm efficiency in a number of applications
(\cite{giro11}, \cite{xifara-etal}). This is the approach adopted
here. We show in the simulation study that, in particular for
stochastic volatility models, we have an efficient algorithm for
estimation and prediction with a lower computational cost, which is
important in practice.

In our SV model this algorithm is then applied following the two steps below.

\begin{enumerate}
\item {\it Sample the latent variables} $\bh$. Assuming the parameters
  as constants, apply (\ref{eq-mala-1}) with $f = f(\by,\bh)$ and
  gradient $\nabla$ calculated with respect to $\bh$. 

\item {\it Sample parameters} $\btheta$. Given $(\by,\bh)$, apply
  (\ref{eq-mmala-1}) and (\ref{eq-mmala-2}) with $f=f(\by,\bh
  |\btheta)f(\btheta)$, gradient $\nabla$ and matrix $\bG$ calculated
  with respect to $\btheta$.
\end{enumerate}

In Appendix \ref{appendix} we provide details on the required expressions of
partial derivatives and metric tensors for both MALA and MMALA. Also,
it is worth mentioning that matrix invertion is less computationally
demanding in the SV model since $G$ has a sparse tridiagonal form.

\subsection{Likelihood and Priors}\label{sec:prior}

The log-likelihood $L_{y|\theta} =  \ln[f(\by,\bh|\btheta)]$ is given by
\begin{eqnarray*}
f(\by,\bh|\beta,\phi,\sigma,\nu) =  f(h_1|\phi,\sigma) \prod_{t=2}^n
f(h_t |h_{t-1},\phi,\sigma)\prod_{t=1}^n f(y_t |h_t,\beta,\nu) 
\end{eqnarray*}
where $h_1 |(\phi,\sigma) \sim N (0,\sigma^2/(1-\phi^2))$, $h_t
|(h_{t-1},\phi,\sigma) \sim N (\phi h_{t-1},\sigma^2)$. In addition,  
\begin{eqnarray*}
f(y_t |h_t,\beta,\nu) = \frac{\nu}{\beta \lambda 2^{1 +
    1/\nu}\Gamma(1/\nu)} \exp \left\{-\frac{h_t}{2} - \frac{1}{2
  \lambda^{\nu}}  
\left|\frac{y_t}{\beta \exp(h_t/2)}   \right|^{\nu}  \right\}
\end{eqnarray*}
for GED errors and 
\begin{eqnarray*}
f(y_t |h_t,\beta,\nu) = \frac{1}{\beta\sqrt{\pi (\nu-2)}}
\frac{\Gamma(\frac{\nu+1}{2})}{\Gamma(\frac{\nu}{2})}  
\left \{1 + \frac{y_t^2}{\beta^2(\nu-2)\exp(h_t)} \right\}^{-(\nu+1)/2} \exp(-h_t/2)
\end{eqnarray*}
for $t$-Student errors.

Following the Bayesian paradigm we need to complete the model
specification with apropriate prior distributions for the parameters.
Independent prior distributions were assigned for $\phi$ and $\sigma$
as in \cite{liu2001} and  
\cite{giro11}, that is $\sigma^2 \sim$ Inv-$\chi^2$(10,0.05),
$(\phi+1)/2 \sim Beta(20,1.5)$. In addition, we propose an Exponential
distribution with mean one as the prior for $\beta$. The prior for the
tail parameter $\nu$ depends on the distribution adopted for the error
terms. For GED errors we
propose the prior for $\nu\sim$ Inv-$\chi^2$(10,0.05) while for
Student-$t$ errors, following \cite{watanabe-asai}, we consider the
truncated exponential density, 
\begin{eqnarray*}
f(\nu) = \lambda \exp\left\{-\lambda(\nu-4)\right\}
\end{eqnarray*}
for $\nu > 4$ and zero otherwise, as the prior for $\nu$. Differently
from \cite{watanabe-asai} we specified $\lambda = 1/3$. 

Denoting the joint prior density of $\btheta$ by $\pi(\btheta)$, the log
prior is then given by,
\begin{eqnarray*}
L_{\theta} = \ln \pi(\btheta) = -\beta - \frac{1}{4\sigma^2} -
11\ln(\sigma) + 19 \ln \left(\frac{1+\phi}{2} \right) 
+ \frac{1}{2} \ln \left(\frac{1-\phi}{2} \right) + \ln f(\nu),
\end{eqnarray*}
where $\ln f(\nu) = - \frac{4}{\nu} - 3\ln(\nu)$ for GED errors and
$\ln f(\nu) = \ln(\lambda) - \lambda(\nu-4)$ for  $t$-Student errors. 

It is worth noting that, in order to employ the algoritms described in
the previous sections, we need to implement a transformation of
$\sigma$, $\phi$ and $\nu$ to the real line. Here we set 
$\sigma = \exp(\gamma)$ and $\phi=\tanh(\alpha)$ as in \cite{giro11},
and we propose  $\nu=\exp(p)$ and
$\nu=\exp(p) + 4$ for GED and $t$-Student errors, respectively.
Of course this introduces Jacobian factors into the acceptance ratios
given by $\frac{d\sigma}{d\gamma} = \exp(\gamma) = \sigma$,
$\frac{d\phi}{d\alpha} = 1 - \tanh^2(\alpha)= 1 - \phi^2$. For GED
errors, $\frac{d\nu}{dp} = \exp(p) = \nu$ and for $t$-Student errors
$\frac{d\nu}{dp} = \nu - 4$.  

\section{Simulations} \label{simulations}

To assess the methodology described in the previous section we
conducted a Monte Carlo study. 
We generated
$m=$1000 replications of 1000 observations from the SV
model (\ref{sv-eqn1})-(\ref{sv-eqn2}) with parameters $\beta=0.65$,
$\phi=0.98$ and two values for $\sigma$, $\sigma \in \{0.05,0.15\}$. These parameter values were used by
\cite{liu2001} and \cite{giro11} among others. We
considered three distributions for the errors: Gaussian, GED with
parameter $\nu=1.6$ and Student's $t$ with $\nu=7$ degrees of
freedom. We then evaluated two estimation schemes: (i) MALA scheme for
both the parameters and the volatilities and (ii) MMALA scheme for the
parameters and MALA scheme for the volatilities (hybrid method). Since
the vector of volatilities has the same dimension as the sample size
(usually thousands of observations) we adopted this hybrid option
instead of using MMALA for both parameters and volatilities. This is
because computation time is relevant in real-life applications.

The true parameter values were used as initial values for the MCMC
samplers and the prior distributions are as described in Section \ref{sec:prior}. 
For each time series we drew 20,000 MCMC samples discarding the first
10,000 samples as a burn-in.

To evaluate the performance of the estimation methods, two criteria
were considered: the bias and square root of the mean 
square error (smse), which are defined as,
\begin{eqnarray}
bias   &=& \frac{1}{m} \sum_{i=1}^m  \hat{\theta}^{(i)} - \theta,\\
smse^2 &=& \frac{1}{m} \sum_{i=1}^m (\hat{\theta}^{(i)} - \theta)^2,
\end{eqnarray}
where $\hat{\theta}^{(i)}$ is the estimate of parameter $\theta$ for
the $i$-th replication, $i=1,\ldots,m$. In this paper we take the
posterior means of $\theta$ as point estimates.

The estimation results are given in Tables 1 and 2. Overall the results are good. 

\begin{center} [ Table 1 around here ] \end{center}
\begin{center} [ Table 2 around here ] \end{center}

\begin{itemize}
\item Gaussian. good results in terms of bias and smse (all parameters). MMALA better excepting fro bias $\beta$

\item GED. good results in terms of bias and smse (all parameters).  MMALA better  for $\beta$ and $\nu$

\item Student's $t$ good results in terms of bias and smse for $\beta$
  and $\nu$ but bad results for $\phi$ and $\sigma$. Maybe we need a
  large sample $n=1000$?  MMALA better excepting for bias $\beta$ 

\end{itemize}

\section{Illustrations} \label{illustrations}

In this section we applied the described methodology to estimate two exchange rate time series data:  
the Pound/Dollar (\pounds/USD) and the Canadian dollar /Dollar (CAN/USD). 
The time series under study are the daily continuously compounded returns in 
percentage, defined as $r_t = 100[\log(P_t)-\log(P_{t-1})]$ where
$P_t$ is the price at time $t$.  
 
The \pounds/USD time series returns covers the period from 1/10/81 to
28/6/85 and the SV model was estimated by \cite{har94}
using quase maximum likelihood methods and by Durbin and Koopman [2001, pp 236] \nocite{durbin-koopman01}
using quase maximum likelihood and Monte Carlo Importance Sampling methods. In both cases the
authors assumed Gaussian errors.

The CAN/USD returns are based on daily noon rates prices. The time
series prices were   obtained from the website http://www.bankofcanada.ca/rates/exchange/
and covers the period from January 2, 2007 to February 7, 2013.

We have 945 and 2509 returns for the \pounds/USD and CAN/USD time
series, respectively. In Figures \ref{PD} and \ref{CD} we show the time series
returns and Table \ref{basic-stats} consigns some descriptive statistics. From this
table, we observe a little skewness and high kurtosis, indicating
asymmetric distributions with heavy tails. In addition, even not
shown, the autocorrelation function indicates non serial correlation.

\begin{center} Figure \ref{PD} around here \end{center}

\begin{center} Figure \ref{CD} around here \end{center}

\begin{center} Table \ref{basic-stats} around here \end{center}

The analysis was done on the demeaned returns. For each time series,
we estimated SV models considering the following three different
distributions for the errors $\varepsilon_t$ in (\ref{sv-eqn1}), the
Gaussian, the GED distribution with parameter $\nu$ and the   
Student's $t$ distribution with $\nu$ degrees of freedom. 

For each time series we drew 150,000 MCMC samples of parameters and
volatilities.  We discarded the first 50,000 as burn-in and skipped
every 25th resulting in a final sample of 4000 values from the
posterior distribution.

The estimated posterior means and standard deviations for each parameter are 
shown in Table \ref{est-results}. We can observe high persistence
estimates ($\phi$). In addition, we  
obtained moderate values of $\nu$ the degrees of freedom in the $t$- Student distribution, 
indicating not too heavy tails\footnote{The maximum likelihood
estimates in \cite{har94} are $\hat{\phi}=0.9912$,
$\hat{\sigma}^2=0.0069$ and $\hat{\gamma}=-0.0879$, then
$\hat{\sigma}=0.0831$ and
$\hat{\beta}=\exp(-\hat{\gamma}/2)=0.9570$. \cite{durbin-koopman01}
report the following maximum likelihood estimates:
$\hat{\phi}=0.9731$, $\hat{\sigma}=0.1726$ and $\hat{\beta}=0.6338$
but do not report the bayesian estimates.}.  
In particular, when comparing point estimates under MALA and MMALA
schemes we note the following.
\begin{itemize}
\item For the \pounds/USD, estimates do not change under Gaussian
  errors but change under GED and Student's $t$ errors with a large
  change in $\nu$ for Student's $t$ errors. The MMALA seems to be more
  efficient to capture heavy tail behaviour.
\item For the CAN/USD, estimates change slightly under Gaussian errors
  but do not change under GED errors. For Student's $t$ errors we
  notice changes in $\beta$ and $\nu$ and again the MMALA scheme
  managed to capture heavy tail behaviour.
\item The posterior standard deviations of $\nu$ are a bit large
corroborating the known fact that this parameter is often difficult to
estimate.
\item The posterior standard deviations of $\beta$ are also large
  for MMALA and Student's $t$ errors.
\end{itemize}

Figure \ref{mcmc} shows the sample autocorrelations, sample paths and
marginal posterior densities of parameters $\beta$, $\sigma$, $\phi$
and $\nu$ for the CAN/USD series using the MMALA sampling scheme under
GED errors. The autocorrelations vanish fairly rapidly and the sample
paths show relatively good mixing in the parameter space.

\begin{center} Figure \ref{mcmc} around here \end{center}

\begin{center} Figure \ref{vol-Pound} around here \end{center}

\begin{center} Figure \ref{vol-Cd} around here \end{center}

In Figures \ref{vol-Pound} and \ref{vol-Cd} are showed the estimated
volatilities $\exp(h_t/2)$ taking the posterior medians of $h_t$ as
point estimates. As
can be seen, the volatilities follow very well the observed volatility
clustering of returns.

The performance of the proposed models and methods can also be
assessed by estimating the Value at Risk (VaR) for multiple time
horizons. From a Bayesian perspective, given the observed values of
returns $\by=\{y_1,\ldots,y_n\}$ point estimates of the one-step ahead
VaR could be obtained using a sample of values drawn from its
predictive distribution, i.e. 
\begin{eqnarray}
E(VaR_{n+1|\by}) \approx \frac{1}{J} \sum_{j=1}^J VaR_{n+1}^{(j)}
\end{eqnarray}
where $VaR_{n+1}^{(j)}$ is the predicted one-step ahead VaR in the
MCMC iteration. Because they are not available analytically we adopt
the following procedure. Given the parameter values and
log-volatilities in the $j$-th iteration we obtain values of
$\{h_{n+1}^{(j)}\}$ by drawing $\eta_{n+1}^{(j)} \sim N(0,
\sigma^{2(j)})$ and setting $h_{n+1}^{(j)} =
\phi^{(j)}h_{n}^{(j)}+\eta_{n+1}^{(j)}$. Next, we generate $L$
replications  $\{\epsilon_{n+1}^{(j,1)},\ldots,\epsilon_{n+1}^{(j,L)}
\}$ 
from the error distribution (with tail parameter $\nu^{(j)}$ for
Student's $t$ or GED distributions). Finally, we form a sample of
returns by setting $y_{n+1}^{(j,k)}= \beta^{(j)}
\exp(h_{n+1}^{(j,k)}/2)\epsilon_{n+1}^{(j,k)}$ which allow us to
approximate $VaR_{n+1}^{(j)}$ of confidence $\alpha$ by the negative
value of the sample $\alpha$-quantile. 

For illustration, we estimated the one day 99\% VaR for the 
last 252 observations (which covers one stock market year
approximately) of both the \pounds/Dollar and the Canadian-Dollar/Dollar time
series. Since we wanted to reproduce a real scenario, the model
parameters were estimated and the VaR calculated based on observations
$y_1,\dots,y_{n-252+i}$, $i=0,\dots,251$. Consequently, we estimated the
model 252 times.

Figure \ref{VaR-PD} shows
the last 252 returns and the VaR estimates using our hybrid MMALA
algorithm for the \pounds/Dollar series. In
252 observations we expected $2.5$ observations below
the VaR. For the Gaussian, Student $t$ and GED distributions we
obtained 8, 7 and 5 observations outside the VaR 
limits, respectively. We note also that the VaR estimates follow very well the
volatility in the market and reacts well to extreme down movements
(large negative return values).

\begin{center} Figure \ref{VaR-PD} around here \end{center}

As for the Canadian-Dollar/Dollar series we note from Figure
\ref{VaR-CD} that, qualitively the results for the Gaussian and GED
errors are better and we 
obtained 2 observations outside the VaR limits in both cases. The VaR's
for Student's $t$ errors on the other hand are quite large
(unnecessarily large from a financial viewpoint). This was indeed
expected given the estimates of $\nu$ in Table 5. The estimate of
$\beta$ is also large compared to Gaussian and GED errors. In our
empirical experience, it is usually better to work with GED
distributions instead of Student's $t$.

\begin{center} Figure \ref{VaR-CD} around here \end{center}

\section{Conclusions}\label{conclusions}

In this paper we discuss a Bayesian estimation of the stochastic
volatility model with Gaussian and two heavy-tailed distributions: GED
and Student's $t$.  Specifically, we implemented the Metropolis
adjusted Langevin (MALA) and Riemann Manifold MALA algorithms. Since
the volatility has dimension equal to the sample size, the
computational time could be high in real-life applications. Then we
implemented a hybrid method: MMALA estimation for the parameters and
MALA for sampling volatilities. These methods were assessed in
simulated data and time series returns. 

As in any Metropolis-Hastings like algorithm, our hybrid sampling
scheme may be sensitive to the choice of the step size parameter
$\epsilon$. Tunning the sampler is simply unavoidable in
practice and we recommend trying two different tuning parameters
during the burn-in period and the stationary phase of the Markov chain
(from which the final sample will be collected).

\section*{Acknowledgements}
This research was partially supported by FAPESP and FAEPEX Grants for
the first author. The third author received support from FAPESP -
Brazil, under grant number 2011/22317-0.

\newpage
\appendix
\section{Appendix}\label{appendix}

In this appendix we present the expressions of gradients and matrix
tensors needed for the implementation of MALA and MMALA for GED and
Student's $t$ errors. For the Gaussian case see \cite{giro11}.  In
what follows, let $\varepsilon_t =  \beta^{-1}\exp(-h_t/2)y_t$. 

\subsection{For GED errors }

\subsubsection*{Sampling volatilities}

The target function is proportional to 
\begin{eqnarray*}
L_{h} =  -\frac{(1-\phi^2)}{2\sigma^2}h_1^2  - \frac{1}{2\sigma^2}
\sum_{t=2}^n (h_t - \phi h_{t-1})^2 - \frac{1}{2}\sum_{t=1}^n h_t
-\frac{1}{2} \sum_{t=1}^n
\left|\frac{\varepsilon_t}{\lambda}\right|^\nu,  
\end{eqnarray*}
therefore the gradient $\nabla_h L_h = \frac{dL_{h}}{dh} = \bs -\br$
where $\bs=(s_1,\ldots, s_n)$ and $\br=(r_1,\ldots, r_n)$ assume values  
\begin{eqnarray*}
s_i  & = & -\frac{1}{2} + \frac{\nu}{4}
\left|\frac{\varepsilon_i}{\lambda}\right|^\nu, \quad i=1,\ldots, n\\ 
r_1 & = & \frac{1}{\sigma^2} (h_1 - \phi h_2), \quad r_n =
\frac{1}{\sigma^2} (h_n - \phi h_{n-1}),\\ 
r_i & = & \frac{1}{\sigma^2} \left[ (h_i - \phi h_{i-1}) -  \phi
  (h_{i+1} - \phi h_{i}) \right], \quad i=2,\ldots, n-1. 
\end{eqnarray*}

\noindent On the other hand the matrix tensor is a symmetric tridiagonal matrix
with  elements $\bG_h(i,j) = -E(\frac{d^2 L_h}{dh_i dh_j})$ for
$i,j=1,\ldots,n$, 
\begin{eqnarray*}
\bG_h(i,i) & = & \frac{\nu}{4} + \frac{1}{\sigma^2}, \quad i=1,n \\
\bG_h(i,i) & = & \frac{\nu}{4} + \frac{1}{\sigma^2}(1+\phi^2), \quad
i=2,\ldots, n-1\\ 
\bG_h(i,i+1) & = & -\frac{\phi}{\sigma^2}, \quad i=1,\ldots, n-1.
\end{eqnarray*}

\subsubsection*{Sampling parameters}

Here  $L_{y|\theta} = \ln[f(\by,\bh|\btheta)]$, i.e.
\begin{eqnarray*}
L_{y|\theta} = 
\frac{1}{2}\ln(1-\phi^2) - n\ln(\sigma) - n\ln(\beta) -\frac{(1-\phi^2)}{2\sigma^2}h_1^2 - 
\frac{1}{2\sigma^2}  \sum_{t=2}^n (h_t - \phi h_{t-1})^2 - \frac{1}{2}
\sum_{t=1}^n \left|\frac{\varepsilon_t}{\lambda}\right|^\nu 
\end{eqnarray*}

\noindent
The partial derivatives of this log-density  with respect to the transformed parameters $(\delta,\gamma,\alpha,p)$ are,
\begin{eqnarray*}
\frac{dL_{y|\theta}}{d\delta} &=& -n + \frac{\nu}{2}\sum_{t=1}^n \left|\frac{\varepsilon_t}{\lambda}\right|^\nu,\\  
\frac{dL_{y|\theta}}{d\gamma} &=& -n + \frac{1}{\sigma^2}(1-\phi^2)h_1^2 + \frac{1}{\sigma^2}\sum_{t=2}^n (h_t - \phi h_{t-1})^2,\\
\frac{dL_{y|\theta}}{d\alpha} &=&  - \phi + \frac{\phi}{\sigma^2}(1-\phi^2)h_1^2 + \frac{(1-\phi^2)}{\sigma^2}\sum_{t=2}^n h_{t-1}(h_t - \phi h_{t-1})\\
\frac{dL_{y|\theta}}{dp} & = & \frac{n}{\nu} \left[\nu -
  \nu\left(\frac{\nu}{\lambda}\frac{d\lambda}{d\nu}\right)  +
  \psi(1/\nu) + \ln(2) \right]-\frac{1}{2} \sum_{t=1}^n
\left|\frac{\varepsilon_t}{\lambda}\right|^\nu \left\{ \ln
\left|\frac{\varepsilon_t}{\lambda}\right|^\nu - \nu
\left(\frac{\nu}{\lambda}\frac{d\lambda}{d\nu}\right) \right\} 
\end{eqnarray*}
where 
\begin{eqnarray*}
\nu \left(\frac{\nu}{\lambda}\frac{d\lambda}{d\nu} \right) =  \ln(2) - \frac{1}{2}\psi(1/\nu) + \frac{3}{2}\psi(3/\nu).
\end{eqnarray*}

\noindent In addition,
\begin{eqnarray*}
E\left(\frac{\partial^2L_{y|\theta}}{\partial\delta^2}\right) & = & -n\nu, \quad
E\left(\frac{\partial^2L_{y|\theta}}{\partial\delta\partial\gamma} \right)=
E\left(\frac{\partial^2L_{y|\theta}}{\partial\delta\partial \alpha} \right) = 0\\
E\left(\frac{\partial^2L_{y|\theta}}{\partial\delta\partial p}\right) 
&=& 
n\left\{1 + \psi(1 + 1/\nu) + \ln(2) -
\nu\left(\frac{\nu}{\lambda}\frac{d\lambda}{d\nu} \right)
\right\}\\ 
E\left( \frac{\partial^2L_{y|\theta}}{\partial \gamma^2} \right) & = &
- 2n, \qquad  E\left( \frac{\partial^2L_{y|\theta}}{\partial \gamma
  \partial \alpha} \right) = -2\phi, \qquad  
E\left( \frac{\partial^2L_{y|\theta}}{\partial \gamma \partial p} \right) = 0 \\
E\left( \frac{\partial^2L_{y|\theta}}{\partial \alpha^2} \right) & = &
- 2\phi^2 - (n-1)(1-\phi^2), \qquad  E\left(
\frac{\partial^2L_{y|\theta}}{\partial \alpha \partial p} \right) =
0\\ 
E\left( \frac{\partial^2L_{y|\theta}}{\partial p^2} \right) & = &
-n\nu \left( \frac{\nu}{\lambda}\frac{d\lambda}{d\nu} \right)^2 +
\frac{n}{\nu} \left\{ (1-1/\nu)\psi_1 (1+1/\nu) 
+ \left[  \psi(1+1/\nu) + \ln(2)\right]^2  \right\}
\end{eqnarray*}
where and $\psi$ and $\psi_1$ are, respectively, the digamma and trigamma functions. 
\vskip .5cm

\noindent Now let $L_{\theta} = \ln \pi(\btheta) = \ln[f(\beta,\phi,\sigma,\nu)]$. Then
\begin{eqnarray*}
\frac{dL_{\theta}}{d\beta} = -1, \quad \frac{dL_{\theta}}{d\gamma} = \frac{1}{2\sigma^2}-11,\quad 
\frac{dL_{\theta}}{d\alpha} = 19(1-\phi)-\frac{1}{2}(1+\phi), \quad  \frac{dL_{\theta}}{dp} = \frac{4}{\nu}-3
\end{eqnarray*}
and the expectations of the second order derivatives of $L_{\theta}$
are given by,
\begin{eqnarray*}
E\left( \frac{\partial^2L_{\theta}}{\partial \gamma^2} \right)  =
-\frac{1}{\sigma^2}, \quad E\left(
\frac{\partial^2L_{\theta}}{\partial \alpha^2} \right) =
-\frac{39}{2}(1-\phi^2), \quad 
E\left( \frac{\partial^2L_{\theta}}{\partial p^2} \right) = -\frac{4}{\nu}. 
\end{eqnarray*}
and zero elsewhere. Finally, we use 
$\nabla_{\theta} \ln f = \frac{dL_{y|\theta}}{d\theta} + \frac{dL_{\theta}}{d\theta}$
and  
$\bG_{\theta} = -E\left(\frac{\partial ^2 L_{y|\theta}}{\partial \theta^2}\right) - E\left(\frac{\partial^2 L_{\theta}}{\partial \theta^2}\right)$.
 
\subsection{For $t$-Student errors}

Next we present those expressions which are different compared with
the GED case.

\subsubsection*{Sampling volatilities}

The target function is proportional to 
\begin{eqnarray*}
L_{h} =  -\frac{(1-\phi^2)}{2\sigma^2}h_1^2  - \frac{1}{2\sigma^2}
\sum_{t=2}^n (h_t - \phi h_{t-1})^2 - \frac{1}{2}\sum_{t=1}^n h_t
-\frac{(\nu+1)}{2} \sum_{t=1}^n  \ln \left(1 +
\frac{\varepsilon_t^2}{\nu-2}\right). 
\end{eqnarray*}

\begin{eqnarray*}
s_i =
-\frac{1}{2} + \frac{1}{2}\frac{(\nu + 1)}{(\nu -2)}\frac{\varepsilon_i^2}{1+\varepsilon_i^2/(\nu-2)}, \quad i=1,\ldots,n
\end{eqnarray*}

\begin{eqnarray*}
\bG_h(i,i) &=& \frac{\nu}{2(\nu+3)} + \frac{1}{\sigma^2}, \quad i=1,n \\
\bG_h(i,i) &=& \frac{\nu}{2(\nu+3)} + \frac{1}{\sigma^2}(1+\phi^2), \quad i=2,\ldots, n-1
\end{eqnarray*}

\subsubsection*{Sampling parameters}

Here  $L_{y|\theta} = \ln[f(\by,\bh|\btheta)]$,
\begin{eqnarray*}
L_{y|\theta} & = & \frac{1}{2}\ln(1-\phi^2) - n\ln(\sigma) - n\ln(\beta) -\frac{(1-\phi^2)}{2\sigma^2}h_1^2  - 
\frac{1}{2\sigma^2}  \sum_{t=2}^n (h_t - \phi h_{t-1})^2  \\
& - & \frac{n}{2}\ln(\nu-2) + n \ln \Gamma \left(\frac{\nu}{2}
+\frac{1}{2}\right) - n \ln \Gamma \left(\frac{\nu}{2}\right)  -
\frac{(\nu+1)}{2} \sum_{t=1}^n  \ln \left(1 +
\frac{\varepsilon_t^2}{\nu-2}\right) 
\end{eqnarray*}

Let $p=\ln(\nu-4)$

\begin{eqnarray*}
\frac{dL_{y|\theta}}{d\beta} 
&=& 
-\frac{n}{\beta} + \frac{\nu+1}{\beta} \sum_{t=1}^n \frac{\varepsilon_t^2/(\nu-2)}{1 + \varepsilon_t^2/(\nu-2)},\\  
\frac{2}{(\nu-4)}\frac{dL_{y|\theta}}{dp}  & = &
n\left[\psi\left(\frac{\nu}{2}+\frac{1}{2}\right)
  -\psi\left(\frac{\nu}{2} \right) - (\nu-2)^{-1}\right] +
\frac{(\nu+1)}{(\nu-2)} \sum_{t=1}^n \frac{\varepsilon_t^2/(\nu-2)}{1
  + \varepsilon_t^2/(\nu-2)}\\ 
& - & \sum_{t=1}^n \ln \left(1 + \varepsilon_t^2/(\nu-2)\right) 
\end{eqnarray*}

\begin{eqnarray*}
E\left(\frac{\partial^2L_{y|\theta}}{\partial\delta^2}\right) &=& -\frac{2n\nu}{\nu+3}\\
E\left(\frac{\partial^2L_{y|\theta}}{\partial\delta\partial p} \right) &=& \frac{-6n(\nu-4)}{(\nu-2)(\nu+1)(\nu+3)}\\
E\left(\frac{\partial^2L_{y|\theta}}{\partial p^2} \right) &=&  \frac{n}{2}\frac{(\nu-4)^2}{(\nu-2)^2} 
\left\{\frac{(\nu-3)(\nu+4)}{(\nu+1)(\nu+3)} + \frac{(\nu-2)^2}{2}
\left[ \psi_1\left(\frac{\nu}{2} + \frac{1}{2}\right) - \psi_1\left(\frac{\nu}{2}\right)\right]  \right\} 
\end{eqnarray*}

Finally, $\frac{dL_{\theta}}{dp} = E\left( \frac{\partial^2L_{\theta}}{\partial p^2} \right) =-\lambda(\nu-4)$.

\end{double}



\clearpage

\begin{table}[htbp]\footnotesize \label{mc-exp1}
\begin{center}
\begin{threeparttable}[b]
\caption{Monte Carlo experiments. Bias and square root of the mean
  squared error of posterior means. Parameters: $\beta=0.65$,
  $\phi=0.98$, $\sigma=0.15$ and $\nu=1.6$ (for GED) and $\nu=7$  (for
  Student's $t$).} 
\begin{tabular}{|llcccccccc|}
\hline 
Errors  & Method &
\multicolumn{2}{c}{$\beta$} &
\multicolumn{2}{c}{$\phi$}  &
\multicolumn{2}{c}{$\sigma$}&
\multicolumn{2}{c}{$\nu$}  \\
        &       &bias    &smse    &bias    &smse    &bias    &smse    &bias   &smse \\
\hline
Gaussian& MALA  &-0.001 & 0.038 &-0.022 & 0.028 & 0.051 & 0.056 &  &  \\ 
        & MMALA & 0.024 & 0.038 &-0.011 & 0.015 & 0.000 & 0.014 &  &  \\ 
GED     & MALA  &-0.002 & 0.032 &-0.042 & 0.051 & 0.090 & 0.099 &-0.011 & 0.128\\ 
        & MMALA & 0.002 & 0.029 &-0.027 & 0.032 & 0.050 & 0.054 & 0.048 & 0.115\\ 
Student's $t$ & MALA & -0.003 & 0.031   &-0.063 & 0.072 & 0.122 & 0.131 & 0.912 & 2.311\\ 
        & MMALA &-0.010 & 0.030 &-0.101 & 0.107 & 0.180 & 0.185 & 0.287 & 1.428\\ 
\hline
\end{tabular}
\end{threeparttable}
\end{center}
\end{table}

\clearpage 

\begin{table}[htbp]\footnotesize \label{mc-exp2}
\begin{center}
\begin{threeparttable}[b]
\caption{Monte Carlo experiments. Bias and square root of the mean
  squared error of posterior means. Parameters: $\beta=0.65$,
  $\phi=0.98$, $\sigma=0.05$ and $\nu=1.6$ (for GED) and $\nu=7$  (for
  Student's $t$).}
\begin{tabular}{|llcccccccc|}
\hline 
Errors  & Method &
\multicolumn{2}{c}{$\beta$} &
\multicolumn{2}{c}{$\phi$}  &
\multicolumn{2}{c}{$\sigma$}&
\multicolumn{2}{c}{$\nu$}  \\
        &       &bias    &smse    &bias    &smse    &bias    &smse    &bias   &smse \\
\hline
Gaussian& MALA  & -0.007 & 0.019  & -0.194 & 0.211 & 0.152 & 0.153 & &  \\ 
        & MMALA & -0.007 & 0.023  & -0.067 & 0.071 & 0.085 & 0.086 & &  \\ 
GED     & MALA  & -0.012 & 0.022  & -0.196 & 0.210 & 0.199 & 0.205 & 0.059 & 0.142\\ 
        & MMALA & -0.013 & 0.025  & -0.107 & 0.112 & 0.132 & 0.133 & 0.109 & 0.152\\ 
Student's $t$ & MALA & -0.014 & 0.027  & -0.193 & 0.205 & 0.231 & 0.238 & 2.163 & 3.145 \\ 
        & MMALA & -0.020 & 0.030  & -0.199 & 0.205 & 0.256 & 0.260 & 1.419 & 2.156\\ 
\hline
\end{tabular}
\end{threeparttable}
\end{center}
\end{table}

\clearpage

\begin{table}[htbp]\footnotesize \label{comp}
\begin{center}
\begin{threeparttable}[b]
\caption{Comparison of methods: MMALA, INLA and MC (Jacquier et
  al.,1994) for Gaussian errors. MMALA and INLA under the same
  conditions. MC used $n=500$ and $\sigma=0.0614$ instead $0.05$ and
  $\sigma=0.166$ instead $0.15$.}
\begin{tabular}{|cclcccc|}
\hline 
\multicolumn{1}{c}{$\phi$} &
\multicolumn{1}{c}{$\sigma$} &
Method & 
\multicolumn{2}{c}{$\hat{\phi}$} &
\multicolumn{2}{c}{$\hat{\sigma}$}\\
 & & & bias    &smse    &bias    &smse \\
\hline
0.98 & 0.15 & MMALA & -0.011 & 0.015 & 0.000 & 0.014\\
     &      & INLA  & -0.011 & 0.017 & 0.575 & 0.586\\
     &      & MC    & -0.010 & 0.020 &-0.064 & 0.080\\
0.98 & 0.05 & MMALA & -0.067 & 0.071 & 0.085 & 0.086\\
     &      & INLA  & -0.074 & 0.120 & 0.238 & 0.245\\
     &      & MC    & -0.070 & 0.127 &-0.079 & 0.099\\
\hline
\end{tabular}
\end{threeparttable}
\end{center}
\end{table}

\clearpage

\begin{table}[h]\footnotesize
\begin{center}
\begin{threeparttable}[b]
\caption{Descriptive Statistics. $n$ is the number of observations} 
\label{basic-stats}
\begin{tabular}{|lccccc|}
\hline 
Time Series & $n$  &Mean    &Std Dev &Skewness &Kurtosis\\ 
\hline
\pounds/USD & 945  &-0.0353 &0.7111  &0.60     & 7.85\\
CAN/USD     & 2509 &-0.0168 &0.6380  &0.14     & 6.18\\
\hline
\end{tabular}
\end{threeparttable}
\end{center}
\end{table}

\clearpage

\setlength{\tabcolsep}{1.2mm}
\begin{table}[h]\footnotesize
\begin{center}
\begin{threeparttable}[b]
\caption{Estimation of stochastic volatility models. Posterior means and standard deviations (in parentheses).}
\label{est-results}
\begin{tabular}{|lllcccc|}
\hline 
Time Series&Method &Errors &$\beta$ &$\phi$ &$\sigma$ &$\nu$\\ 
\hline
\pounds/USD&MALA & Gaussian      & 0.6156 (0.0115) & 0.9824 (0.0042) & 0.0903 (0.0016) &                \\
           &     & GED           & 0.3351 (0.0056) & 0.9980 (0.0008) & 0.0904 (0.0014) & 2.0572 (0.1118)\\
           &     & Student's $t$ & 0.6353 (0.0136) & 0.9827 (0.0043) & 0.0841 (0.0015) &10.5511 (1.8144)\\
           &MMALA& Gaussian      & 0.6311 (0.0146) & 0.9847 (0.0052) & 0.0752 (0.0017) &                \\
           &     & GED           & 0.6095 (0.0157) & 0.9920 (0.0036) & 0.0665 (0.0015) & 1.6538 (0.1047)\\
           &     & Student's $t$ & 0.9832 (0.2201) & 0.9875 (0.0046) & 0.0584 (0.0013) & 4.7574 (0.4305)\\
CAN/USD    &MALA & Gaussian      & 0.5524 (0.0066) & 0.9873 (0.0022) & 0.0812 (0.0009) &                \\
           &     & GED           & 0.5546 (0.0069) & 0.9875 (0.0022) & 0.0839 (0.0009) & 1.7670 (0.0590)\\
           &     & Student's $t$ & 0.5699 (0.0071) & 0.9905 (0.0019) & 0.0606 (0.0006) &12.6578 (2.0301)\\
           &MMALA& Gaussian      & 0.5579 (0.0079) & 0.9921 (0.0023) & 0.0628 (0.0009) &                \\
           &     & GED           & 0.5701 (0.0089) & 0.9853 (0.0032) & 0.0815 (0.0011) & 1.7311 (0.0777)\\
           &     & Student's $t$ & 0.8182 (0.1503) & 0.9895 (0.0027) & 0.0631 (0.0027) & 5.1043 (0.9192)\\
\hline
\end{tabular}
\end{threeparttable}
\end{center}
\end{table}

\clearpage


\begin{figure}[h]\centering
\includegraphics[scale=0.85,keepaspectratio=true]{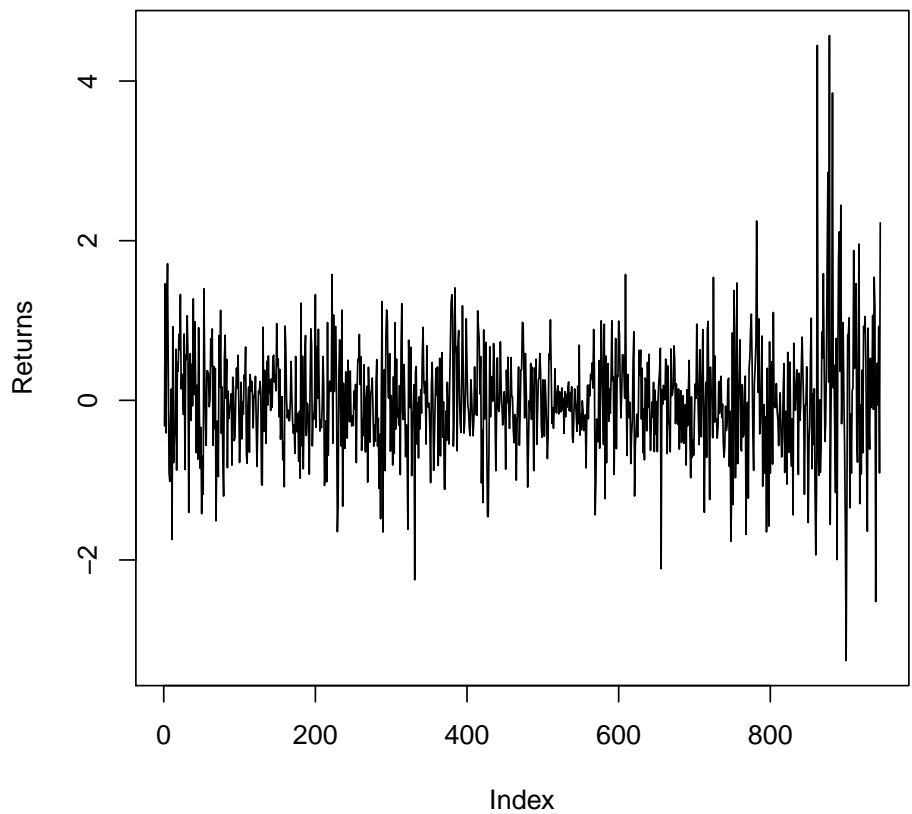}
\caption{Pound/Dollar time series returns.}
\label{PD}
\end{figure}

\clearpage


\begin{figure}[h]\centering
\includegraphics{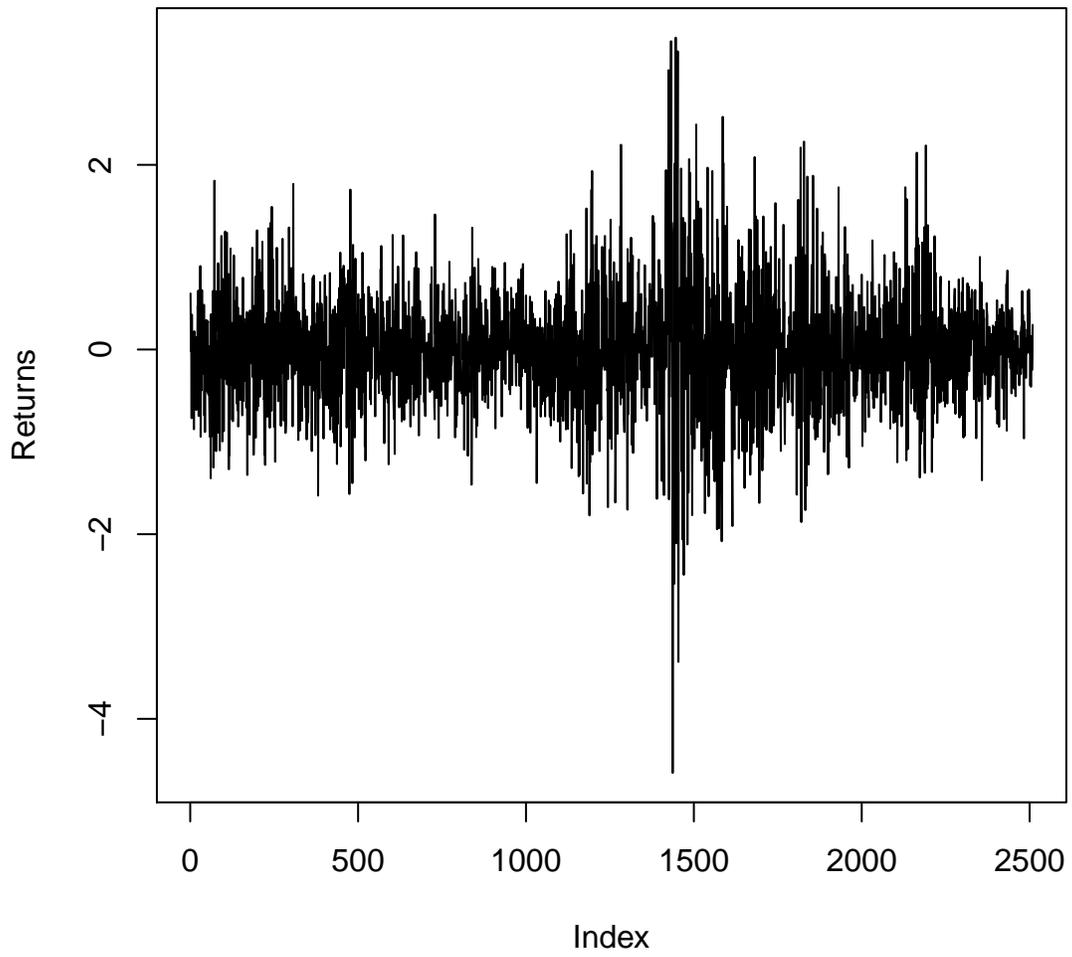}
\caption{Canadian dollar/Dollar time series returns.}
\label{CD}
\end{figure}

\clearpage

\begin{figure}[h]\centering
\end{figure}

\setkeys{Gin}{width=\textwidth,height=4.5in}
\begin{figure}[h]\centering
\includegraphics{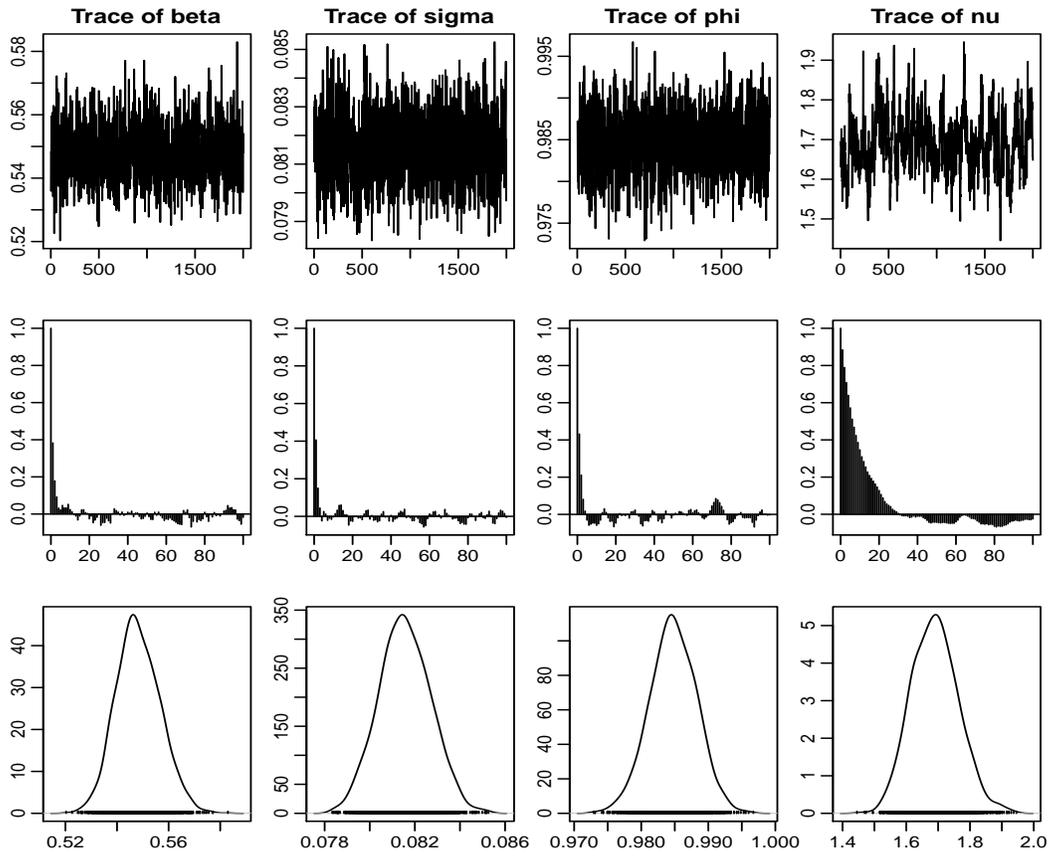}
\caption{Sample autocorrelations, sample paths and marginal posterior
  densitieso for the CAN/USD series using the MMALA sampling scheme under
  GED errors.}
\label{mcmc}
\end{figure}

\clearpage

\begin{figure}[h]\centering
\includegraphics{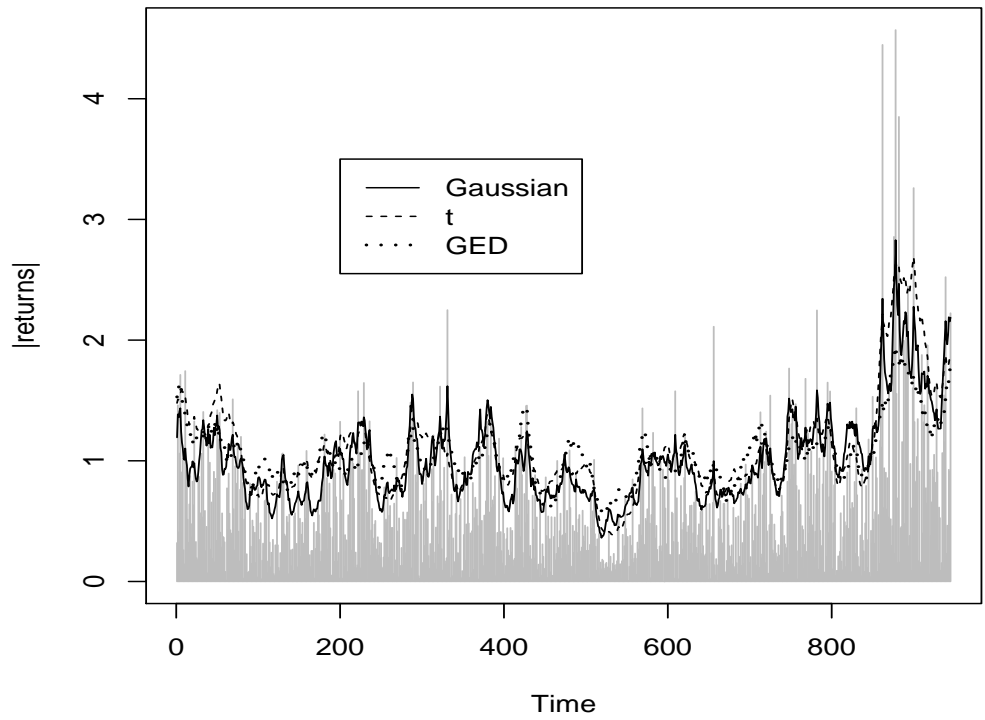}
\caption{Absolute returns for the Pound/Dollar series and estimated
  volatilities using MMALA under the three different errors.}
\label{vol-Pound}
\end{figure}

\clearpage

\begin{figure}[h]\centering
\includegraphics{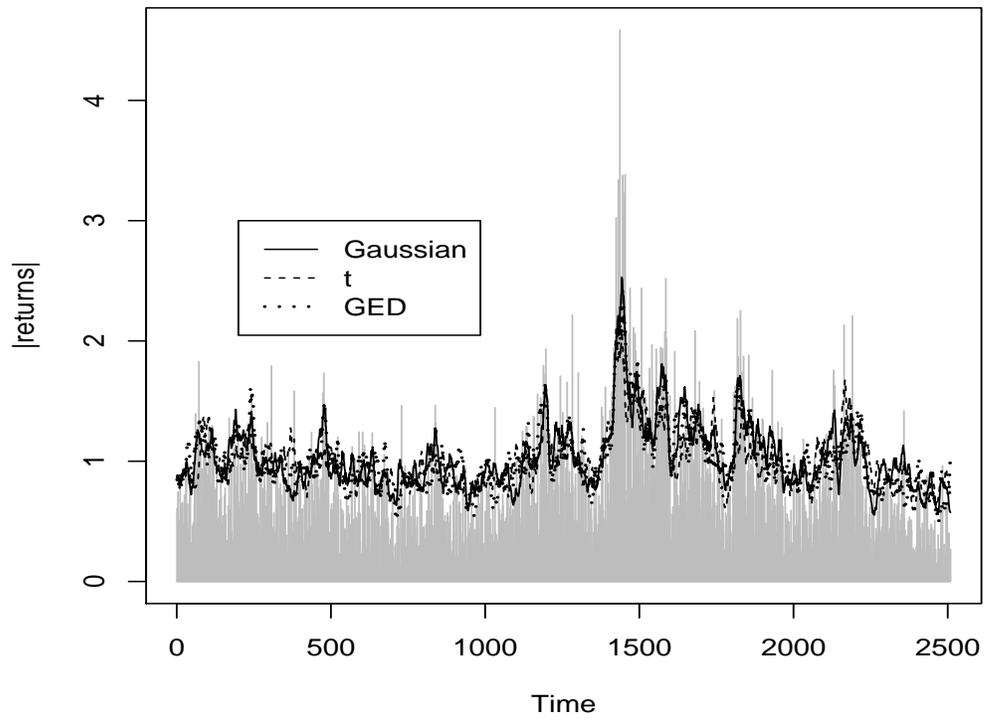}
\caption{Absolute returns for the Canadian Dollar/Dollar series and estimated
  volatilities using MMALA under the three different errors.}
\label{vol-Cd}
\end{figure}

\clearpage


\begin{figure}[h]\centering
\includegraphics{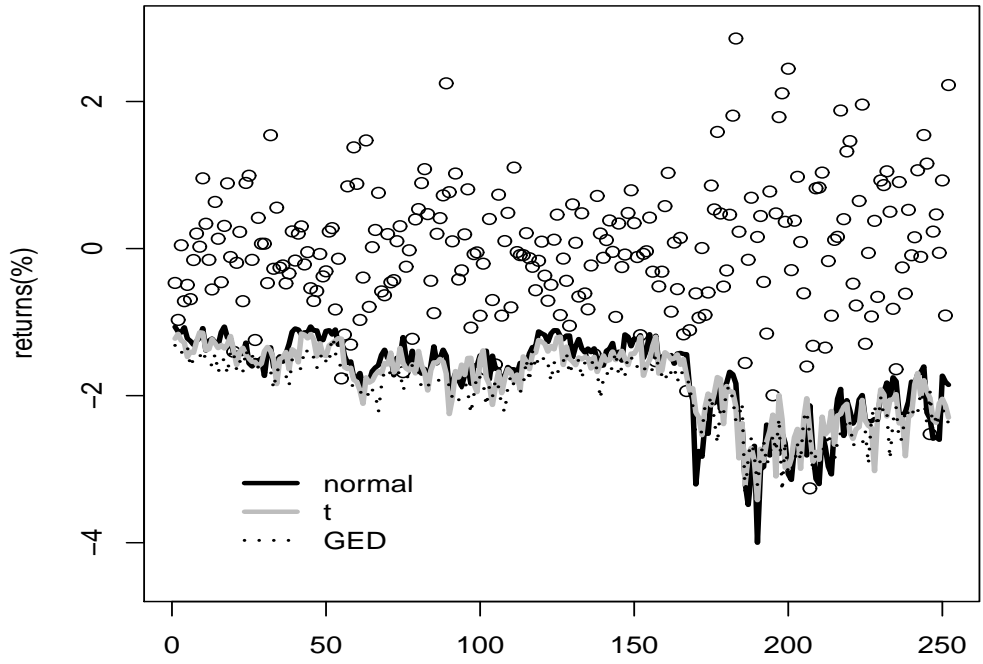}
\caption{99\% Value at risk of Pound/Dollar exchange rates using the
  MMALA scheme.}
\label{VaR-PD}
\end{figure}


\clearpage

\begin{figure}[h]\centering
\includegraphics{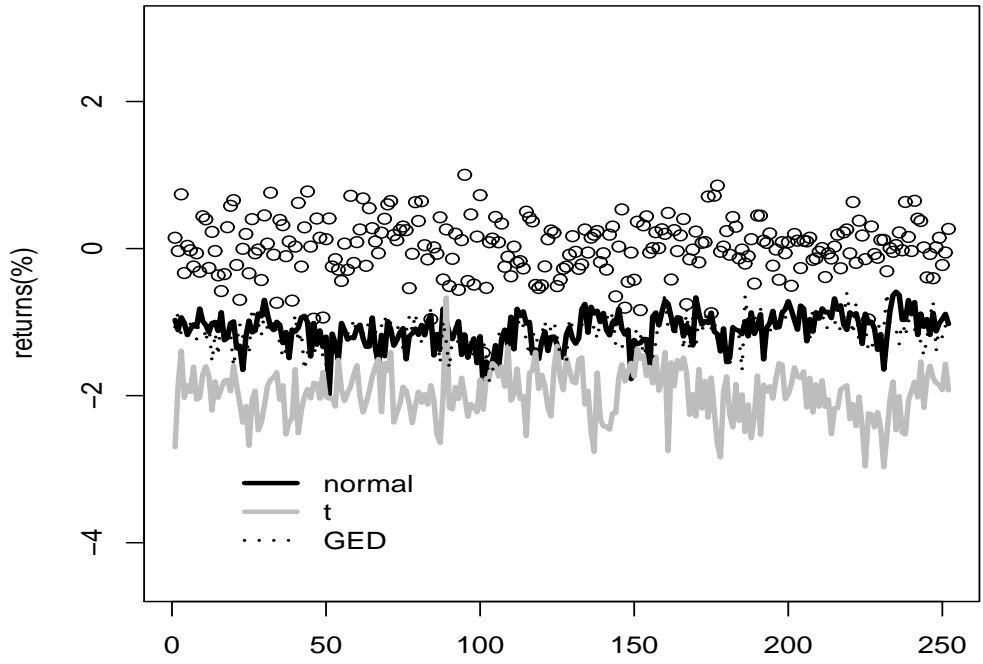}
\caption{99\% Value at risk of Canadian-Dollar/Dollar exchange rates using the
  MMALA scheme.}
\label{VaR-CD}
\end{figure}


\end{document}